**Title: Roadmap for systematic identification and analysis of multiple biases in causal inference**

**Authors:** Wijesuriya R.[1,2*], Hughes R.A.[3,4], Carlin J.B[1,2], Peters R.L [5,6]., Koplin J.J [6,7]., and Moreno-Betancur M. [1,2]

**ORCiD IDs:** Wijesuriya R. (0000-0003-1023-4065), Hughes R.A. (0000-0003-0766-1410), Carlin J.B(0000-0002-2694-9463), Peters R.L (0000-0002-2411-6628), Koplin J.J (0000-0002-7576-5142), Moreno-Betancur M. (0000-0002-8818-3125)

Affiliations:

1. Clinical Epidemiology & Biostatistics Unit (CEBU), Murdoch Children's Research Institute
2. Clinical Epidemiology & Biostatistics Unit (CEBU), Department of Paediatrics, University of Melbourne
3. MRC Integrative Epidemiology Unit, University of Bristol
4. Population Health Sciences, Bristol Medical School, University of Bristol
5. Department of Paediatrics, University of Melbourne
6. Murdoch Children's Research Institute
7. Child Health Research Centre, University of Queensland

*Correspondence: r.wijesuriya@unimelb.edu.au

**Funding:** Prof Moreno-Betancur is supported by funding from the Australian National Health and Medical Research Council (NHMRC) Investigator Grant 2009572. The Murdoch Children's Research Institute is supported by the Victorian Government's Operational Infrastructure support program. Dr Hughes is supported by a Sir Henry Dale Fellowship that is jointly funded by the Wellcome Trust and the Royal Society (grant 215408/Z/19/Z).

**Conflicts of interest:** The authors have no conflicts of interest to disclose.

**Data availability statement:** All software code for the simulation study and case study analysis can be found here: [https://github.com/rushwije/MBA_roadmap]. Case study data cannot be made publicly available due to ethics requirements but are available from study investigators upon reasonable request. Please direct any case study data requests to health.nuts@mcri.edu.au.

**Title: Roadmap for systematic identification and analysis of multiple biases in causal inference**

## Abstract

Observational studies examining causal effects rely on unverifiable assumptions, the violation of which can induce multiple biases. Quantitative bias analysis (QBA) methods examine the sensitivity of findings to such violations, generally, by producing estimates under alternative assumptions, incorporating external information. Although substantial guidance exists for implementing QBA, there is limited guidance on how to systematically determine the assumptions underlying a primary causal analysis and the potential violations that should guide bias analysis. Consequently, many assumptions remain implicit, leading to selective and therefore misleading QBA. To address this gap, we propose a roadmap for systematically identifying and analysing multiple biases. Briefly, this consists of (i) articulating the assumptions underlying the primary analysis through specification and emulation of the "ideal trial" that defines the causal estimand and depicting these assumptions using a causal diagram; (ii) extending the diagram to depict alternative assumptions under which biases may arise; (iii) obtaining a single estimate that simultaneously corrects for all potential biases. We illustrate the roadmap using an investigation of the effect of breastfeeding on risk of childhood asthma, and through simulations illustrate the need for analysing multiple biases jointly rather than one at a time.

## Abbreviations:

ACE, Average Causal Effect; CB, Confounding Bias; MB, Measurement Bias; DAG, Directed Acyclic Graph; FN, False Negative; FP, False Positive; OR Odds Ratio; PBA, Probabilistic Bias analysis; QBA, Quantitative Bias Analysis; SB, Selection Bias; SE, Standard Error; Simulation Interval; TN, True Negative; TP, True Positive.

Many epidemiological studies aim to quantify the causal effect of an exposure on an outcome, generally relying on causal assumptions such as exchangeability, consistency and positivity. Violations of causal assumptions can lead to systematic causal bias — a discrepancy between the expected value of the causal effect estimate and the true (unknown) value, versions of which are often categorised as confounding, measurement (or information) and selection bias. [1] For example, confounding and selection bias may arise as a violation of exchangeability assumptions. [2]

Quantitative bias analysis (QBA) examines the sensitivity of findings to violations of the assumptions made in the primary analysis, generally by producing estimates under alternative assumptions.[1] Specifically, QBA focuses on alternative assumptions under which the observed data alone do not permit unbiased estimation of the causal effect. Therefore, QBA relies on analytic approaches that incorporate external information to address these biases.

Although substantial guidance has emerged for conducting QBA— including analytic approaches and best practice recommendations[1,3-5]— there is currently limited guidance on how to systematically identify the assumptions underlying a primary causal analysis and the potential violations of those assumptions (i.e. likely sources of systematic bias) in order to design the bias analysis. Given the lack of a structured approach, in many cases the primary analysis assumptions remain implicit, which results in selective identification and consideration of biases in QBA, undermining the transparency and completeness of bias assessments.

In addition, even when multiple potential biases are identified, a common strategy for QBA in practice involves analysing each one at a time (producing multiple alternative estimates).[1,3] Given that different bias sources do not necessarily act independently of each other or in the same direction, this strategy can lead to misleading conclusions about the overall impact of biases.[1,5] Alternatively, multiple bias modelling approaches can be used, which jointly account for dependencies between different sources of bias. However, this can be a complex undertaking as it requires specifying model(s) for multiple biases and their parameters simultaneously. [3] To set-up appropriate bias models, it is essential to understand all plausible assumption violations and the overall causal structure of the problem.

In this work we propose a roadmap for systematically identifying and analysing multiple biases (see Box 1) and illustrate the roadmap using a case study investigating the effect of breastfeeding on risk of childhood asthma. Further, using simulations based on this case study—where multiple dependent biases are plausible—we compare the commonly used one-at-a-time bias analysis with simultaneous bias analysis (a type of multiple bias modelling), to illustrate the importance of using multiple biases jointly and motivate the use of the proposed roadmap.

## Preliminaries: Illustrative case study example and target estimand

We consider a case study based on a published investigation [6] from the HealthNuts cohort study, a prospective observational study of 5279 infants aged 11-15 months recruited via immunization clinics in Melbourne, Australia from 2007-2011 and followed up at ages 4, 6 and 10 years.[7] Here we focus on the specific question: *To what extent does receiving any (versus no) breastfeeding in the first 12 months reduce the risk of asthma at 6 years?* This question is of interest because some organizations recommend that any breastfeeding is better than none. [8-10] In this context, "breastfeeding" is generally interpreted as feeding breast milk to the infant by any mode (i.e., including bottle-fed pumped breast milk) and is therefore how we interpret it in this manuscript.[8,11]

We next proceed to define the target estimand mathematically. Later we will utilise the concept of an ideal trial as an alternative way of defining it. Let $A$ denote a binary indicator for any breastfeeding in the first 12 months ($A = 1$) or not ($A = 0$), and let $Y$ denote the binary outcome, with $Y = 1$ if child is diagnosed with asthma by age 6, and $Y = 0$ otherwise. Although $A$ summarises a time-varying behaviour over 12 months, under the plausible assumption that breastfeeding is either established in the first one to two weeks or is never established, $A$ is equivalent to an indicator of breastfeeding initiation in those first weeks [12] [13] and the breastfeeding status of a child does not change thereafter (e.g. if breastfeeding was not initiated, then the child will never be breastfed, and vice versa). This allows us to consider "any breastfeeding" as a (single time-) point exposure, which is the setting of focus in the paper.

Our target estimand is the average causal effect (ACE) of any breastfeeding in the population of all neonates born in Melbourne who may be able to be breastfed, specifically the causal risk ratio $\mathrm{P}(Y^1 = 1)/\Pr(Y^0 = 1)$, where $Y^a$ is the *potential* (*a.k.a. counterfactual*) outcome when intervening to set the exposure $A$ to the value $a$.

We note that there are multiple versions of treatment in this setting as there are many ways (versions) in which the recommendation that any breastfeeding is better than none (the treatment) can be implemented by families (e.g. the duration and mode of any breastfeeding). Still, as it has been shown[14], in the absence of missing data and measurement error, the target estimand is identifiable under specific causal consistency, positivity and conditional exchangeability assumptions. Specifically, positivity and exchangeability need to consider a set of confounders $\boldsymbol{C}$ that includes variables that may affect both treatment and the version of treatment. Under these assumptions, the causal risk ratio is identifiable from observable data based on the g-formula: $P(Y^a = 1) = \sum_{\boldsymbol{c}} P(Y = 1|A = a, \boldsymbol{C} = \boldsymbol{c}) \times P(\boldsymbol{C} = \boldsymbol{c})$.[15]

This implies that unbiased estimation is possible using a number of methods[15]. We focus on outcome regression, given its widespread use.[16] The outcome regression models have mean specification as follows:

$$\log[\Pr(Y = 1|A = a, \boldsymbol{C} = \boldsymbol{c})] = \beta_0^{RR} + \beta_1^{RR} A + \boldsymbol{\beta_2^{RR}}'\boldsymbol{C} \qquad (1)$$

where $e^{\beta_1^{RR}}$ equal the causal risk ratio, if the outcome model in (1) is correctly specified and there is no effect-measure modification across strata of $\boldsymbol{C}$, (in addition to the above-mentioned assumptions).

## The proposed roadmap for multiple bias analysis

### Step 1: Systematic determination of all assumptions in the primary analysis and depicting these using DAGs

To facilitate this step, we propose defining the ideal randomized trial which, as we have argued elsewhere,[17] defines the causal effect of ultimate interest in any causal study. This approach offers an alternative to the potential outcomes framework for precisely specifying the relevant estimand.[17] As such, the ideal trial is the referent for consideration of biases.

#### *Using the ideal trial to articulate assumptions for the primary analysis*

The ideal trial is a hypothetical trial with an infinite and representative sample, perfect adherence and follow-up, and no missing data or measurement error. It is defined by specifying its protocol components: eligibility criteria, treatment strategies, assignment procedures, follow-up period, and outcome measure.[18] The observational analysis can then be seen as an attempt to emulate the ideal trial. The emulation of each component will rarely be perfect, with systematic bias avoided only under unverifiable identifiability assumptions. By considering the emulation of each protocol component in turn, we can determine all the identifiability assumptions and assess their plausibility.

#### *Directed acyclic graphs*

Once the identifying assumptions are systematically recognised, it is helpful to depict these using a DAG (where possible, as not all assumptions can be depicted in DAGs).[19] Comprehensive reviews and tutorials on DAGs can be found elsewhere.[15,20] Briefly DAGs are graphs with nodes representing variables (and their measurements thereof) related to the research question of interest, (e.g., treatments, outcome, covariates, missingness indicators) connected by arrows, which depict assumed causal relationships between variables. The primary analysis of a study typically assumes that under the assumptions depicted in the corresponding DAG (and any additional assumptions that are not amenable to depiction in a DAG), standard (non-QBA) analytic approaches (e.g. (1)) suffice to obtain unbiased estimates from the observable data.

#### *Application to the case study*

We begin by detailing features of the published analysis that are relevant for examining potential biases, along with notation for related variables. Firstly, the HealthNuts cohort and thus the published study only included children with consent (denoted by $S$; 1: yes, 0:no) from a guardian who was able to speak and understand English (denoted by $E$; 1: yes, 0:no). Second, the study relied on parental reports of breastfeeding ($A^*$), and asthma ($Y^*$), as more objective measures (e.g. doctor diagnosis) were not collected by the investigators. Third, a potential confounder, pre-exposure hypertension ($U$; 1: presence, 0: absence), was not measured.[21] Fourth, there was substantial loss to follow-up, leading to missing data in $Y^*$ (denoted by response indicator $R_y$; where $R_y = 1$ if $Y^*$ observed and $R_y = 0$ otherwise).

In Table 1 we outline the protocol components for what we conceive as the ideal trial capturing the estimand of ultimate interest in the published study (column 1), as well as their emulation, as implicit in the published primary analysis (column 2) and the corresponding identifiability assumptions (column 3).

*Table 1

Figure 1 below shows the assumed (simplified) DAG under the primary analysis. An extended DAG that shows the assumed relationships between all measured confounders can be found in the published investigation.[6]

*Figure 1

**Step 2: Considering plausible violations of assumptions of the primary analysis and using DAGs to identify potential sources of bias**

The second step of the roadmap involves considering plausible violations of the assumptions in the primary analysis and depicting these graphically using DAGs. Specifically, we are interested in violations that arise due to variables that are unobserved, incomplete and/or mismeasured in the study (including unobserved data for those not selected into the study sample). In practice, one might focus on the plausible violations that are most likely to influence the results, as we illustrate with our example. DAGs are useful in bias analysis,[1] as they facilitate an understanding of how biases may enter an analysis if the assumptions in the primary DAG are violated, particularly confounding and selection bias, as described further below (see Table 2 and Figure 2).

***Causal biases and their graphical structures***

Causal biases can be broadly categorised into three types, each stemming from distinct processes: selection, confounding and measurement bias (see Table 2). The different biases can be depicted using DAGs, generally as arising from open non-causal pathways from A to Y, except in the case of measurement bias and selection bias due to effect measure modification[22] as explained in detail in Table 2. Nonetheless, using DAGs to depict assumptions about the measurement error, along with other assumption violations, can help inform multiple bias model specification in Step 3 of the roadmap. In Figure 2 we provide a few example DAGs for different types of bias, although different DAGs are possible.

*Table 2

*Figure 2

***Application to the case study***

Below we consider the plausible violations of the assumptions made in the primary analysis as identified in Table 1 and Figure 1 and hypothesise their structure by proposing an alternative DAG, as shown in Figure 3, where we consider the full structure under the alternative assumptions (shown in grey).

Figure 3*

*Selection bias*

The primary analysis assumes that infants recruited to HealthNuts in 2007-2011 are representative of all infants aged 1 year born in Melbourne in 2006 -2010 who are able to be breastfed (see *eligibility criteria* in Table 1). Given the high immunization rates in Victoria during this period[23], and the low infant mortality[24], we assume differences between the study sample and the target population due to these factors are minimal and resulting selection bias (if any) is negligible, so we do not consider possible violations in the alternative DAG.[25] Additionally, the primary analysis did not exclude infants who may not be able to be breastfed (subgroup 1: infant unable to ingest or digest breastmilk; subgroup 2: mother unable to breastfeed), as data on these factors were not available. Therefore, there is a possible positivity violation in the emulation (subgroup 1), with a form of Type II selection bias possible due to extrapolation if the effect in these subgroups is not defined (subgroup 1) or different (subgroup 2). However, the size of subgroup 1 is expected to be small and we do not expect a different effect in subgroup 2, so these biases are expected to be negligible, and we do not consider them in the alternative DAG.

On the other hand, the study was restricted to infants with consent from a guardian, $S = 1$, who can speak/understand English, $E = 1$, but the primary analysis assumed no effect measure modification by these variables (Figure 1). However, parents who consented were more likely to have infants with a history of eczema, which is a potential effect measure modifier (i.e. the effect of breastfeeding on asthma could be different among those with eczema and no eczema).[6,7] This implies a likelihood of effect measure modification by proxy for $S$, in which case the primary analysis estimate is subject to Type II selection bias (Table 2). If $E$ (which is a strong proxy for ethnicity) is an effect measure modifier, a similar bias can arise. Figure 3 thus depicts the plausibility of $E$ and $S$ being effect measure modifiers in the risk ratio scale.

The primary analysis assumed that missingness in the outcome was only dependent on the observed variables, $\boldsymbol{C}$ and $A^*$. Figure 3 considers potential collider-stratification (selection) bias due to missingness in the outcome being additionally dependent on the outcome, which is plausible through unmeasured common causes of the outcome and missingness in the outcome such as healthcare-seeking behaviours.

*Confounding bias*

The primary analysis adjusted for a set of measured confounders. Figure 3 depicts a potential residual confounding due to unmeasured gestational hypertension (*U*)[21], a common cause of breastfeeding and child asthma. Note that there may be other potential unmeasured confounders that could be explored, but we focus on gestational hypertension for illustration.

*Measurement error*

The primary analysis assumes perfect measurement of exposure and outcome. Figure 3 depicts potential measurement error due to parent reports of the outcome and exposure. Although adjusting for $\boldsymbol{C}$ in the primary analysis excludes the possibility of bias due differential measurement error of an additive form (as defined in Table 2) and dependent measurement error, this does not entirely preclude measurement bias. [26] In general, measurement error will lead to bias, except when the errors are non-differential and independent, and there is no causal effect of $A$ on $Y$.[26,27]

**Step 3: Conducting quantitative multiple bias analysis**

When there are multiple potential sources of bias, the most common analytic strategy is to analyse the biases one at time. However, this can result in misleading results because it implicitly assumes the sources of bias act independently. Instead, a multiple bias modelling approach[28] should be used.

One multiple bias modelling strategy is to analyse biases serially (often using single bias modelling approaches[1]), in the reverse order in which the biases are assumed to have occurred.[1] However, the final estimate from this strategy can vary depending on the order chosen[1,4,29], which is problematic when a plausible order is unclear. A more flexible multiple bias modelling strategy analyses the biases simultaneously without requiring an a priori assumption about their order.[30,31] Therefore, for step 3 of the roadmap, we adopt the simultaneous strategy, specifically using a recent approach that we review in the next section. [31] Simulations have shown that this approach produces estimates of the causal odds ratio with negligible bias, [31] and performs similarly to serial analysis when the latter is conducted using the "true" order of biases, under relatively simple scenarios (e.g. measurement error in exposure dependent only on the true exposure, selection dependent only on exposure and outcome etc.). [30,31]

Regardless of the strategy used, all bias analysis modelling approaches require bias parameters,[28,32] that stipulate the strength and direction of the association between observed and unobserved/mismeasured/incomplete variables under alternative assumptions. Different bias parameters are required for the different types of bias (e.g., see Table S2). By their nature, bias parameters cannot be estimated directly from observed data and therefore need to be informed by external data sources and expert opinion, substantive literature, or internal/external validation data.[3] Investigators may treat bias parameters as fixed (simple bias analysis), vary them systematically across a range of plausible values (multidimensional bias analysis), or conduct a probabilistic bias analysis (PBA) in which a (joint) prior probability distribution is assigned to the bias parameter(s) followed by a Bayesian analysis or utilisation of a Monte Carlo sampling procedure.[1,3] PBA allows incorporating the investigators' uncertainty about the bias parameters in the final estimate, provided the uncertainty is properly captured by the distribution used. In the context of a multiple bias analysis, a PBA may be necessary, as with multiple biases, there will be numerous combinations of fixed bias parameters to consider which can make it difficult to summarize and report results clearly.[3]

***Simultaneous multiple bias modelling approach***

The approach considered works by expressing the desired joint probability (or model), which involves unobserved/mismeaured/incomplete variables, as the product of the joint probability of the observed variables and a bias-adjusting probability that depends on bias parameters according to the assumed alternative assumptions. This enables derivation of an expression for the bias-adjusting probability that is then used to generate values for the full (ideal) data under the alternative assumptions and thereby obtain estimates under alternative assumptions.[31] Below we describe the steps involved in a setting like the case study, where Figure 3 depicts the alternative assumptions (in grey):

I. ***Derive an expression for the bias-adjusting probability and posit bias models***

In a setting like the case study, the desired joint probability is $\mathrm{P}(Y = y, A = a, \boldsymbol{C} = \boldsymbol{c}, U = u)$ and the observed joint probability is $\mathrm{P}(Y^* = y^*, A^* = a^*, C = c | S = 1, E = 1, R_y = 1)$. Let $\pi$ denote the bias-adjusting probability such that the desired joint probability is the product of $\pi$ and the observed joint probability. That is:

$$\pi = \frac{\mathrm{P}(Y = y, A = a, \boldsymbol{C} = \boldsymbol{c}, U = u)}{\mathrm{P}(Y^* = y^*, A^* = a^*, C = c | S = 1, E = 1, R_y = 1)} \qquad (2)$$

Based on the assumptions encoded in Figure 3, $\pi$ can be expressed as follows (see supplementary files for detailed steps):

$$\pi = \frac{\mathrm{P}(U = u | A, Y) \times \mathrm{P}(Y = y | A, Y^*, \boldsymbol{C}) \times \mathrm{P}(A = a | A^*, Y^*, \boldsymbol{C})}{\mathrm{P}(S = 1 | A^*, Y^*) \times \mathrm{P}(E = 1 | A^*, Y^*) \times \mathrm{P}(R_y = 1 | A^*, Y^*, \boldsymbol{C})}$$

Parametric models are then posited for each of the factors in this expression, referred to as the bias models. With binary variables, logistic regression models can be used as follows:

$$\begin{aligned}
\mathrm{logit}\big(\mathrm{P}(A = 1 | A^*, Y^*, \boldsymbol{C})\big) &= \gamma_0 + \gamma_1 A^* + \gamma_2 Y^* + \boldsymbol{\gamma_3^T C} \\
\mathrm{logit}\big(\mathrm{P}(Y = 1 | Y^*, A, \boldsymbol{C})\big) &= \alpha_0 + \alpha_1 Y^* + \alpha_2 A + \boldsymbol{\alpha_3^T} C \\
\mathrm{logit}\big(\mathrm{P}(U = 1 | A, Y)\big) &= \delta_0 + \delta_1 A + \delta_2 Y \\
\mathrm{logit}\big(\mathrm{P}(S = 1 | A^*, Y^*)\big) &= \theta_0 + \theta_1 A^* + \theta_2 Y^* + \theta_3 A^* Y^* \\
\mathrm{logit}\big(\mathrm{P}(E = 1 | A^*, Y^*)\big) &= \lambda_0 + \lambda_1 A^* + \lambda_2 Y^* + \lambda_3 A^* Y^* \\
\mathrm{logit}\Big(\mathrm{P}\big(R_y = 1 | A^*, Y^*, \boldsymbol{C}\big)\Big) &= \eta_0 + \eta_1 A^* + \eta_2 Y^* + \boldsymbol{\eta_3^T C}
\end{aligned} \qquad (3)$$

In the above model(s) for $S$ (and $E$), an interaction between $A^*$ and $Y^*$ is included to allow for the association between $S$ (and $E$) and $Y^*$ to vary across levels of $A^*$ and vice versa, reflecting the potential effect measure modification of the association between $A$ and $Y$ by $S$ (and $E$). The regression coefficients of bias models are bias parameters and as such have to be assigned values or distributions that are elicited from external information.

### *II. Impute the unobserved variables and derive selection weights for each individual*

The unobserved/incomplete/mismeasured variables $U, A, Y,$ are singly imputed for each record based on the corresponding bias models in (3) using the elicited values for the bias parameters and the observed data. In our example, with binary $U, A$ and $Y$, the imputations are created by sampling from a Bernoulli trial for each patient with success probability set to the predicted probability from the corresponding bias model.

Selection probabilities are estimated from the corresponding bias models for $E$ and $R_y$ using the elicited values for the bias parameters and the observed data. These can be multiplied if assumed independent as in Figure 3, and then used to derive inverse probability weights for each record in the analytic sample: $\widehat{W} = 1/[\hat{P}(S = 1|A^*, Y^*) \times \hat{P}(E = 1|A^*, Y^*) \times \hat{P}(R_y = 1|A^*, Y^*, \boldsymbol{C})]$. These weights are used in the next step to create a pseudo-population that is representative of the target population.

### *III. Perform weighted outcome regression*

The bias-adjusted estimates of the causal risk ratio under the alternative assumptions can be obtained by fitting the outcome regression models in (1) to the imputed data using weighting.

The uncertainty in the bias parameters can be incorporated in the final estimate using PBA. First, bias parameters are sampled from a plausible distribution over a number of replications, then Steps 2 and 3 are repeated for each draw, producing multiple bias-adjusted effect estimates. A simulation interval (SI) can be calculated using the range of estimates (e.g., the 2.5$^{th}$ and the 97.5$^{th}$ percentiles as the 95% SI)[3]. To also incorporate sampling variability into the SI, the bias-adjustment procedure can be replicated over bootstrap samples or via a normal approximation as detailed elsewhere.[28,31]

### *Application to the case study*

We used the above analytic method to analyse the multiple biases identified in the case study simultaneously (considering the assumptions depicted in Figure 3), and for comparison we used the same approach to examine each bias separately, i.e. one at a time, ignoring other biases. We used a probabilistic approach with 1000 simulations. Where available, the key bias parameter distributions were obtained via a literature search and cross-checked with substantive context experts; otherwise, they were elicited from experts [see Table S2]. Other bias parameters were set to approximations based on the observed data. Sampling variability was incorporated by repeating the procedure over 1000 bootstraps.

Figure 4 shows results for the bias analyses alongside those for the primary analysis. Although the results across approaches were only minimally different, the simultaneous multiple bias analysis resulted in somewhat stronger causal effect estimates, suggesting a larger reduction in risk of asthma under any breastfeeding versus not, compared with analysing the biases separately, one at a time, or ignoring the potential biases, as in the primary analysis.

*Figure 4

## Simultaneous vs one-at-a-time analysis of multiple biases: simulation study

### *Simulation study aims and design*

We conducted a proof-of-concept simulation study to compare two strategies for analysing multiple biases — simultaneous analysis of multiple biases and the more common strategy of one-at-a-time bias analysis —to demonstrate the need for analysing the multiple biases together.

Data were generated and analysed as detailed in Table 3. The key bias parameters considered are detailed in Table 4. Note that in the case study, there were two plausible generalizability biases (Type II selection biases): due to restriction to those with parents/guardians who can speak and understand English, and due to restriction to those with parent/guardian consent). However, in the simulation study, for brevity, we only consider the former generalizability bias.

*Table 3
*Table 4

### *Simulation study results*

As shown in Table 5, the simultaneous analysis of biases resulted in approximately unbiased estimates of the (log) risk ratio (<10% relative bias) with nominal coverage (~ 0.93-0.96). The primary analysis and all one-at-a-time bias-analyses produced estimates with comparatively large relative biases (>20%), and some under- or overcoverage. Most one-at-a-time analysis produced bias-eliminated coverage closer to nominal levels, indicating that the poor coverage observed was largely due to bias in point estimates.

*Table 5

## Discussion

In this paper we introduced a three-step roadmap for facilitating the systematic identification and simultaneous analysis of multiple potential sources of bias, illustrated using a real case study. The steps include identifying all assumptions in the primary analysis through specification and emulation of the ideal trial and depicting these using DAGs where possible, [17], considering plausible assumption violations and depicting these in alternative DAGs to identify the resulting causal biases, and conducting a multiple bias analysis considering the biases simultaneously to obtain a single estimate assessing robustness to the multiple biases. Further, proof-of-concept simulations illustrated how one-at-a-time analyses of biases can lead to estimates of the causal effect with large relative biases while a simultaneous analysis can produce approximately unbiased casual effect estimates with appropriate coverage, given correctly specified bias model and bias parameters. The roadmap thus supports more robust QBA, particularly in settings where multiple potential sources of bias are present, encouraging a more comprehensive and transparent assessment of biases. Nevertheless, as with any bias analysis or indeed any causal inference, implementing the roadmap is not without challenges, as we discuss next.

The ideal trial in the first step of the roadmap is defined as the hypothetical, perfect randomized experiment that would answer the ultimate scientific question of interest.[17] As we discuss elsewhere,[17] in specifying it, a balance must be struck between relevance to decision-making and feasibility of estimation with the available data. For instance, the outcome measure may be selected as the gold standard measure that would be used in practice, even in the absence of this measure in the observational study. This is different to a practice that is common in studies using the related "target trial framework", which tend to specify the target trial (also a hypothetical trial) such that it is closely aligned with features of the observational data.[33] As we have argued, focusing on the ideal trial that considers relevance as well as feasibility of estimation is necessary for examining biases relative to the ultimate, practically-relevant estimand.[17] However we recognize that striking this balance is not always straightforward. In our example, we considered these trade-offs when specifying the ideal trial.

Although DAGs can be a useful tool in bias analysis and play a prominent role in our roadmap, as we have mentioned earlier and detailed in Table 2, not all biases can be identified (e.g. measurement bias) and/or demonstrated (e.g. selection bias due to effect measure modification) using DAGs. Nevertheless, representing the causal assumptions graphically as far as possible can help to clarify how the variables—and their sometimes-imperfect measurements—that induce bias are related. In some cases, researchers may have incomplete knowledge of the full data-generating mechanism, making it challenging to develop a DAG. Even in such situations, DAGs remain valuable for explicitly laying out assumptions—however uncertain—making the causal structure and alternative scenarios more transparent and clarifying the potential pathways through which biases may arise.

Another obvious caveat in modelling multiple biases, as we illustrate in the third step of the roadmap, is the requirement of specifying distributions for many bias parameters. Our case study illustrates how investigators can make a reasonable attempt to source and justify bias parameter distributions [see Table S2]. However, further research on approaches for calibrating marginal parameters obtained from external sources to conditional bias parameters, as available for other sensitivity analysis methods,[34] are required to enable wider application of multiple bias modelling approaches. In addition, there is often uncertainty around the true values of bias parameters, making misspecification likely if fixed values are assumed. As a result, it is always good practice to incorporate the uncertainty in bias parameters as explained either using multidimensional bias analysis or PBA. We have illustrated the latter in our case study application.

All studies estimating causal effects either explicitly or implicitly make causal assumptions that, if violated, may result in biases with respect to the estimand of interest. Investigators should move beyond simply making qualitative statements about the possible biases in their studies and quantitatively demonstrate the robustness of their results to multiple possible biases simultaneously. This would allow greater transparency to readers and journal editors in identifying areas for improvement and inform future research efforts. The roadmap presented, incorporating the ideal trial specification and DAGs, offers a systematic guide to the necessary steps for achieving a more rigorous, comprehensive, and transparent bias analysis.

**Tables and figures**

**Step 1**: Articulate all identifiability assumptions made in the primary analysis by specifying the ideal trial (i.e. the trial that defines the causal effect of interest) and considering the assumptions required to emulate it, depicting these assumptions using directed acyclic graphs (DAGs) where possible.

**Step 2:** Consider plausible violations of the assumptions in the primary analysis due to variables that are unobserved/incomplete/mismeasured in the study (e.g. unmeasured confounding, gold standard measures, etc.) and depict these alternative scenarios using DAGs to assess possible sources of causal biases in the primary analysis.

**Step 3**: Obtain a single estimate assessing robustness to all possible sources of bias identified by applying a simultaneous multiple bias analysis approach.

Box 1: Roadmap for quantitative bias analysis considering multiple biases simultaneously

**Table 1:** Ideal trial specification for systematically identifying potential causal biases in the breastfeeding and asthma case study

| Protocol component | Ideal trial | Emulation (implicit in published analysis[6]) | Identifiability assumptions |
|---|---|---|---|
| Eligibility Criteria | ***Target population:***<br>All neonates born in Melbourne who may be able to be breastfed | ***Analytic sample selection:***<br>All HealthNuts participants, who were infants aged 11-15 months in Melbourne in 2007-2011, recruited during routine 12-month immunization sessions across Melbourne, with parents/guardians who consented to participate in the study and could read and understand English. Inclusion was regardless of missingness in outcome, which was handled via inverse probability weighting (IPW) with weights estimated conditional on the selected confounders and measured exposure*<br><br>Given lack of data, no exclusions made on the basis of infant not being able to be breastfed either due to (i) infant being unable to ingest or digest breastmilk or (ii) mother unable to breastfeed | The HealthNuts study participants are representative of (i.e. not systematically different from) the population of all infants born in Melbourne in 2006-2010 who may be able to be breastfed<br><br>Missingness in the outcome is independent of outcome given the measured exposure and the selected confounders<br><br>There are no HealthNuts participants who may not be able to be breastfed due to condition (i) detailed in the previous column of the table, and the causal effect of interest is not different (i.e. no effect modification) for infants who may not be able to be breastfed due to condition (ii). |
| Treatment strategies | ***Treatment arms in the trial:***<br>(1) Never breastfeed in the first 12 months<br>(2) Breastfeed at least once in the first 12 months | ***Treatment/exposure measure:***<br>Classified into treatment arm consistent with parent report of ever breastfeeding in the first 12 months collected when infant aged 11-15 months | Parents can accurately recall and report breastfeeding directly or via bottle during the first year of life at 11-15 months |
| Assignment procedures | Randomisation at recruitment without blind assignment | Adjustment for the following set of measured pre-birth confounders: child's sex measured at birth, maternal age at birth, maternal smoking during pregnancy, parental ethnicity, SES (socioeconomic status), maternal asthma, paternal asthma, family history of allergy, number of siblings | The set of variables (or more specifically, their measures) represents a sufficient set of confounders to eliminate all confounding |
| Follow-up period | Start: Birth<br>End: Child aged 6 years | Start: Child aged 11-15 months<br>End: Child aged 6 years | Time zero of follow-up and the timing at which eligibility was assessed are later than in the ideal trial and misaligned with the start of treatment (birth). This could lead to (i) *misclassification bias* and/or (ii) *selection bias.*[35]<br><br>i. *Misclassification bias*: We assume that breastfeeding is either established in the first weeks or is never established, so the treatment strategies in the ideal trial are equivalent to "initiate breastfeeding" vs "do not initiate breastfeeding". Thus, under this assumption, the treatment strategies in the ideal trial are clearly distinguishable at time zero, and perfect adherence can be assumed, meaning they can be treated as time-invariant (i.e. as point treatments). Additionally, an indicator of ever breastfed in the first 12 months in the observational study can validly emulate the treatment assignment indicator in the ideal trial, thereby avoiding potential misclassification bias.<br><br>ii. *Selection bias*: This is avoided under the assumptions already noted under "Eligibility criteria" (i.e. the HealthNuts participants recruited at age 11-15 months are representative of all infants born in Melbourne who are able to be breastfed). |
| Outcome | ***Outcome measure*** Clinically diagnosed asthma at age 6 (i.e., physician assessment of clinical history, with assessment at age 6, in conjunction with lung function testing demonstrating expiratory airflow limitation) | ***Outcome measure*** Parental report of doctor-diagnosed asthma and either wheezing episodes in the last 12 months or use of common asthma medication (relievers and preventers) in the previous 12 months when child aged 6 years | Parent self-reports are accurate measures of clinically diagnosed asthma at age 6 |

* Although the published study had missing data in exposure and confounders, here we assume them to be fully observed for brevity, as only a small proportion of data were missing (exposure: 2.4%; confounders: 13.2%). The resulting study sample size was n=4579

**Table 2:** Description and graphical structure of key causal biases

| Causal bias | Description | Graphical structure |
|---|---|---|
| Selection bias | Occurs when the sample used for analysis (i.e., analytic sample) is not representative of the population for which inference is of interest (i.e. target population).<br>This can occur either because (i) the population included in the study (i.e. study sample) is not representative of the target population or, (ii) because the analytic sample (the observed portion of the study sample) is not representative of the study sample, or (iii) both [13]. | Broadly can arise in two ways:<br><br>***Collider stratification bias/ Type I selection bias***<br>Formally in a DAG the structure of this bias can be represented as arising from a non-causal (biasing) path that becomes open after restricting the analysis or conditioning on one (or more) level(s) of a common effect of two variables (known as a "collider"), one of which is either the exposure or a cause of the exposure, and the other is either the outcome or a cause of the outcome.[36,37] In Figure 2(i), conditioning on $R_y$ opens a biasing pathway between $A$ and $Y$ via $R_y$.<br><br>***Generalizability bias/Type II selection bias***<br>This type of bias arises due to restricting to one (or more) level(s) of an effect measure modifier (a variable such that the magnitude of the causal effect differs across its strata). The causal structure of effect modification cannot be denoted as a biasing pathway but may be depicted as shown in Figure 2(ii) where $E$ is an effect measure modifier of the causal effect of $A$ on $Y$ for illustration. Other depictions are also possible as illustrated elsewhere. [22,38] This type of bias is scale dependent (multiplicative vs. additive) as it arises due to effect measure modification. [22,39] |
| Confounding bias | Occurs due to uncontrolled common causes of the exposure and outcome. | Formally in a DAG the structure of this bias can be depicted as resulting from an open non-causal (biasing) "backdoor" path between the exposure and the outcome. Figure 2(iii) shows an example of confounding bias, where there is an open backdoor path between $A$ and $Y$ via $U$. |
| Measurement bias | Occurs due to discrepancy between the measured value of a quantity and its true value as determined by what is considered to be the gold standard instrument. This includes misclassification in the case of categorical data. | The structure of this bias can be denoted in a DAG by depicting both the measured variable and the underlying true variable (e.g., $A$ and $A^*$). The specific structure of the measurement error varies depending on whether the errors are assumed to be differential or non-differential and/or dependent or independent (as defined below) and cannot always be represented as a biasing pathway. [26] And there may be measurement bias even in the context of non-differential, independent measurement error.<br><br>Measurement errors are said to be independent if the measured outcome and the exposure are statistically independent of one another conditional on the true outcome and the true exposure (i.e. $Y^* \perp$ |

| | |
|---|---|
| | $A^*\|Y, A)$ and dependent otherwise. Measurement errors are said to be differential if the measured outcome (or measured exposure) depends not only on its own true value, but itself or this latter relationship also depend on the true exposure (or true outcome), and non-differential otherwise. Specifically, two forms of differential error can be distinguised[40]: (i) additive form, where $A$ (or $Y$) leads to a systematic shift in the $Y^*$ (or $A^*$ ), and (ii) multiplicative form, where the $A$ (or $Y$) modifies the strength of the relationship between $Y - Y^*$ (or $A - A^*$). As such (ii) cannot be depicted in a DAG.<br><br>Figure 2 (iv) provides an example of non-differential and dependent (via $\boldsymbol{C}$) measurement errors. |

**Table 3:** Proof-of-concept simulation study data design (see Table S1 for an extended table with additional details on the parameter values)

| | **Details** |
|---|---|
| **Data generation** | We simulated 1000 datasets (each of size n=2000) using the parametric models detailed below, based on Figure 3 in the main text. The non-bias model parameters were set to values estimated by fitting analogous models to the case study data (ignoring the unobserved variables). The values of the bias parameters ($pU$, $OR_{A-U}, OR_{Y-U}, pE\ , OR_{(Y-A)E}, OR_{Y-R_y}, OR_{Y^*-Y}, OR_{A^*-A}, OR_{Y-R_y}$) were derived based on a literature review and cross-checking with substantive context experts (see Table S2 for relevant references). For simplicity, in the data generation under each scenario, we considered the bias parameters to be fixed.<br><br>1. Observed confounders $\boldsymbol{C}$ were generated sequentially based on the hypothesized causal relationships structure in the published analysis. See [6] for more details.<br><br>2. Next, the unmeasured confounder, $U$, breastfeeding exposure $A$, Ability to speak/understand English $E$, and the outcome $Y$ were generated using the following models,<br>$$U \sim Bernoulli\left(logit^{-1}(pU)\right),$$<br>$$A \sim Bernoulli\left(logit^{-1}\left(\omega_0 + \boldsymbol{\omega}_1^T\boldsymbol{C} + log(OR_{A-U})\,\mathrm{U}\right)\right),$$<br>$$\mathrm{E} \sim Bernoulli\left(logit^{-1}(pE)\right),$$<br>$$\mathrm{Y} \sim Bernoulli\left(logit^{-1}(\phi_0 + \log(\alpha)\,\mathrm{A} + \phi_1^{\mathrm{T}}\boldsymbol{C} + \log\,(OR_{Y-U})\mathrm{U} + 0.18\mathrm{E} + \log\left(OR_{(Y-A)E}\right)\mathrm{A} \times \mathrm{E})\right),$$<br>where $\alpha$ was set to a fixed value determined via trial and error to achieve ~ 80% power for estimating the ACE using the target analysis models at the conventional 0.05 significance level in all scenarios (before introducing any missing values). Based on the above criterion, $\alpha = (-0.43)$<br><br>3. Next, misclassified exposure and outcome were generated:<br>$$A^* \sim Bernoulli(logit^{-1}\left(\pi_0 + \boldsymbol{\pi}_1^T\boldsymbol{C} + \log\,(OR_{A^*-A})A)\right)$$ |
| | $$Y^* \sim Bernoulli(logit^{-1}(\lambda_0 + \boldsymbol{\lambda_1^T C} + \log(OR_{Y^*-Y})\,Y))$$<br>4. Finally, missingness in the outcome was generated:<br>$$R_Y \sim Bernoulli(logit^{-1}\left(\tau_0 + \tau_1 A + \boldsymbol{\tau_2^T C} + \log\left(OR_{Y-R_y}\right)Y)\right)$$ |
| **Analysis** | Following data generation, the one-at-a-time and simultaneous bias analysis strategies were applied to each dataset using the analytic approach described in section "Multiple bias modelling approach" to estimate for the ACE of any breastfeeding on asthma at age 6 in the log risk ratio scale (i.e., $\log(\beta_{RR})$). |
| **Performance assessment** | The "true" values of the target parameter in the log scale (i.e. ACE in log risk ratio scale) was obtained by fitting model (1) to a very large sample without selection and all the unobserved data. These were used to compute the following performance measures:<br><br>Bias: Average difference between the estimate from bias analysis and the true value<br>Relative bias: Bias relative to the true value as a percentage<br>Empirical SE: Standard deviation of the estimate from the bias analysis<br>Model-based SE: Average of the estimated standard error of the estimate from the bias analysis<br>Coverage probability: Proportion of replications where the estimated 95% confidence interval contained the true value<br>Bias-eliminated coverage: Proportion of replications in which the estimated 95% CI contained the mean value of the estimate from bias analysis (across replications) |

Note $logit^{-1}((.)) = \exp(.)/(1+\exp(.))$

**Table 4:** Bias parameters values used in the data generating models in the simulation study

| Bias | Parameter * | Parameter value |
|---|---|---|
| Selection bias (Collider stratification bias) | Relative change in odds of asthma at age 6 in those whose guardian responded at age 6 vs not, $OR_{Y-R_y}$ | 1.20 |
| Selection bias (Generalizability) | Probability of being able to speak and understand English ($p_E$ ) (marginal) | 0.85 |
| | Relative change in OR of asthma-breastfeeding in those who can speak/understand English compared to those who cannot, $OR_{(Y-A)E}$ | 0.70 |
| Confounding bias due to unmeasured confounder gestational hypertension, $U$ | Prevalence of gestational hypertension ($p_U$) (marginal) | 0.10 |
| | Relative change in odds of being ever breast-fed for infants with maternal gestational hypertension vs those without, $OR_{A-U}$ | 0.60 |
| | Relative change in odds of having asthma at age 6 for infants with maternal gestational hypertension vs those without, $OR_{Y-U}$ | 1.30 |
| Measurement bias-exposure | Relative change in odds of parent-reporting as ever breast fed in those ever-breast-fed vs never, $OR_{A^*-A} = [\mathrm{TP}/\mathrm{FN}]/[\mathrm{FP}/\mathrm{TN}]$<br><br>Where:<br>$\mathrm{TP} = \mathrm{p_A} \times \mathrm{N} \times sens$<br>$\mathrm{FN} = (\mathrm{p_A} \times \mathrm{N}) - \mathrm{TP}$<br>$\mathrm{FP} = (1 - \mathrm{p_A}) \times \mathrm{N} \times (1 - spec)$<br>$\mathrm{TN} = ((1 - \mathrm{p_A}) \times \mathrm{N}) - \mathrm{FP}$<br><br>$\mathrm{p_A}$=0.94 | 1.70<br>$sens$=0.90<br>$spec$=0.84 |
| Measurement bias-outcome | Relative change in odds of parent reporting as child having asthma in those with asthma vs not, $OR_{Y^*-Y} = [\mathrm{TP}/\mathrm{FN}]/[\mathrm{FP}/\mathrm{TN}]$<br><br>Where TP, FN, FP and TN as above, calculated using $\mathrm{p_Y}$=0.10 | 0.54<br>$sens$=0.83<br>$spec$=0.90 |

OR: Odds ratio, TP: true positives, FN: false negatives, FP: false positives, TN: true negatives

*Parameters are conditional unless stated otherwise and were often approximated using marginal values from the literature - see Table S2 for relevant references.

**Table 5** Comparative performance of the bias analysis strategies in estimating the log risk ratio, $\log(\beta_{RR})$= log (0.48)

| | **Method** | **Bias** | **Relative bias** | **Emp SE** | **Model SE** | **Coverage** | **Bias eliminated Coverage** |
|---|---|---|---|---|---|---|---|
| | Primary analysis | 0.30 | -50.62 | 0.50 | 0.65 | 0.96 | 0.92 |
| One-at-a-time bias analysis | CB only | 0.31 | -50.90 | 0.50 | 0.65 | 0.95 | 0.94 |
| | MB-A only | 0.34 | -56.16 | 0.23 | 0.23 | 0.74 | 0.95 |
| | MB-Y only | 3.17 | -527.04 | 6.04 | 0.73 | 0.69 | 0.00 |
| | SB-collider stratification only | 0.31 | -51.35 | 0.50 | 0.65 | 0.97 | 0.94 |
| | SB-generalizability only | 0.28 | -47.10 | 0.50 | 0.65 | 0.97 | 0.94 |
| Multiple bias analysis | All biases | -0.03 | 5.25 | 0.32 | 0.33 | 0.94 | 0.96 |

CB: Confounding bias, MB: measurement bias, SB: Selection bias, SE: Standard error

## Supplementary files

Bias-adjusting probability derivation

The desired joint probability refers to the joint probability of all variables that are required to perform the desired analysis while also encoding the alternative assumptions considered for the sensitivity analysis, i.e. $Y^*, A^*, \boldsymbol{C}, Y, A, U$, as well as any selection variables that have been conditioned on in obtaining the observed data. The observed joint probability refers to the joint probability of the observed variables $(Y^*, A^*, \boldsymbol{C})$ in the selected study sample (i.e. in the stratum defined by $S = 1, E = 1$ and $R_y = 1$)

$$\text{bias-adjusting weight} = \frac{\text{desired joint probability}}{\text{observed joint probability}}$$

$$= \frac{P(Y = y, A = a, U = u, A^* = a^*, Y^* = y^*, \boldsymbol{C} = \boldsymbol{c})}{P(Y^* = y^*, A^* = a^*, C = c | S = 1, E = 1, R_y = 1)}$$

$$= \frac{P(Y = y, A = a, U = u | A^* = a^*, Y^* = y^*, \boldsymbol{C} = \boldsymbol{c}) \times P(A^* = a^*, Y^* = y^*, \boldsymbol{C} = \boldsymbol{c})}{P(Y^* = y^*, A^* = a^*, C = c | S = 1, E = 1, R_y = 1)}$$

$$= \frac{P(Y = y, A = a, U = u | A^* = a^*, Y^* = y^*, \boldsymbol{C} = \boldsymbol{c}) \times P(A^* = a^*, Y^* = y^*, \boldsymbol{C} = \boldsymbol{c})}{\frac{P(S = 1, E = 1, R_y = 1 | A^* = a^*, Y^* = y^*, \boldsymbol{C} = \boldsymbol{c}) \times P(A^* = a^*, Y^* = y^*, \boldsymbol{C} = \boldsymbol{c})}{P(S = 1, E = 1, R_y = 1)}}$$

$$\propto \frac{P(Y = y, A = a, U = u | A^* = a^*, Y^* = y^*, \boldsymbol{C} = \boldsymbol{c})}{P(S = 1, E = 1, R_y = 1 | A^* = a^*, Y^* = y^*, \boldsymbol{C} = \boldsymbol{c})}$$

$$= \frac{P(U = u | A = a, Y = y, A^* = a^*, Y^* = y^*, \boldsymbol{C} = \boldsymbol{c}) \times P(Y = y | A = a, A^* = a^*, Y^* = y^*, \boldsymbol{C} = \boldsymbol{c}) \times P(A = a | A^* = a^*, Y^* = y^*, \boldsymbol{C} = \boldsymbol{c})}{P(S = 1, E = 1, R_y = 1 | A^* = a^*, Y^* = y^*, \boldsymbol{C} = \boldsymbol{c})}$$

$$= \frac{P(U = u | A = a, Y = y) \times P(Y = y | A = a, Y^* = y^*, \boldsymbol{C} = \boldsymbol{c}) \times P(A = a | A^* = a^*, Y^* = y^*, \boldsymbol{C} = \boldsymbol{c})}{P(S = 1, E = 1, R_y = 1 | A^* = a^*, Y^* = y^*, \boldsymbol{C} = \boldsymbol{c})}$$

... where the last step follows from the conditional independence assumptions encoded in Figure 3.

Table S1: Simulation study design using the ADEMP structure [1]

| | **Details**<br>*Recommendations from Table 1 of Morris et al. (Stat Med. 2019;38(11):2074-2102.), with additions* |
|---|---|
| **Aims** | 1. Assess the performance of the simultaneous bias -adjustment approach in estimating the causal risk ratio and difference under different strengths of bias and varying degree of misspecification of bias parameters.<br>*2.* Compare the performance of the simultaneous and one-at-a-time bias-adjustment strategies |
| **Data generating mechanisms** | Data were simulated from parametric models, based on Figure 3 in the main text with model parameters estimated by fitting analogous models (ignoring the unobserved variables) as explained below<br><br>Sample size (N): 2000<br><br>*Observed confounders*<br>Observed confounders were generated sequentially as follows based on the hypothesized causal relationships structure in the published analysis. See [2] for more details :<br>1. First the following confounders were generated independently by randomly assigning a % of participants to categories based on observed proportions in each category in the real data.<br>• family history of allergic disease ($\mathrm{FHx}$): 31% No, 69% Yes<br>• paternal asthma($\mathrm{FPa}$): 86% No,14% Yes<br>• maternal asthma($\mathrm{FMa}$): 85% No,15% Yes<br>• parental ethnicity($\mathrm{PEth}$): 60% Australian,10% Asian, 30% Other<br>• number of older siblings($\mathrm{NSibs}$): 50% None, 33% One, 13% Two, 4% Three<br>• child's sex at birth($\mathrm{Sex}$): 51% Male,49% Female<br>• SES($\mathrm{SES}$): 20% Quintile 1, 20% Quintile 2, 21% Quintile 3, 19% Quintile 4,20% Quintile 5<br><br>2. Maternal age at childbirth, environmental smoking, and maternal smoking were generated conditional on SES using a linear and logistic regression models as follows:<br><br>$$\mathrm{MAge} \sim N(30.82 + 2.20\mathrm{SES}_2 + 2.14\mathrm{SES}_3 + 2.65\mathrm{SES}_4 + 3.75\mathrm{SES}_5, 4.8)$$<br>$$\mathrm{MSmk} \sim Bernoulli(logit^{-1}(-2.10 - 1.25\mathrm{SES}_2 - 0.84\mathrm{SES}_3 - 1.48\mathrm{SES}_4 - 2.57\mathrm{SES}_5))$$ |
| | where $\mathrm{SES}_2, \mathrm{SES}_3, \mathrm{SES}_4, \mathrm{SES}_5$ denotes indicator variables for $SES$ quintiles 2-5<br><br>3. Gestational age was generated using a logistic regression model conditioning on maternal age at childbirth, SES, and maternal smoking:<br><br>$$\mathrm{GAge} \sim Bernoulli(logit^{-1}(-2.51 + 0.01\mathrm{MAge} + 0.04\mathrm{SES}_2 - 0.16\mathrm{SES}_3 - 0.31\mathrm{SES}_4 - 0.16\mathrm{SES}_5 - 0.08\mathrm{MSmk}))$$<br><br>4. Mode of delivery was generated using a logistic regression model conditioning on SES and maternal age at childbirth:<br>$$\mathrm{DMode} \sim Bernoulli(logit^{-1}(-2.92 + 0.07\mathrm{MAge} - 0.05\mathrm{SES}_2 - 0.07\mathrm{SES}_3 - 0.16\mathrm{SES}_4 - 0.11\mathrm{SES}_5))$$<br><br>5. Low birthweight was generated conditioning on mode of delivery, gestational age, SES, maternal smoking, maternal age at childbirth and env smoking:<br><br>$$\mathrm{LBW} \sim Bernoulli(logit^{-1}(-4.81 + 0.70\mathrm{DMode} + 3.85\mathrm{GAge} + 0.01\mathrm{MAge} + 0.14\mathrm{SES}_2 - 0.01\mathrm{SES}_3 + 0.16\mathrm{SES}_4 + 0.34\mathrm{SE} + 0.66\mathrm{MSmk}))$$<br><br>Parameters in the models 1-5 above were closely based on associations observed in the real case study data [2]<br><br>*Unmeasured confounder*<br><br>6. Gestational hypertension was generated independently from a binomial distribution using a success probability sourced from the literature (see Table 3 in main text) :<br>$$U \sim Bernoulli(logit^{-1}(p_U))$$<br><br>*Exposure (true unobserved)*<br><br>7. Breast feeding was generated using a logistic regression model conditioning on all observed confounders and unmeasured confounder as below: |
| | $A\sim Bernoulli(logit^{-1}(1.79-0.17\mathrm{Sex}-0.56\mathrm{NSibs}_2-0.99\mathrm{NSibs}_3-1.06\mathrm{NSibs}_4-0.38\mathrm{LBW}+0.31\mathrm{PEth}_2+0.36\mathrm{PEth}_3+0.04\mathrm{FMa}-0.11\mathrm{FPa}-0.01\mathrm{FHx}-0.10\mathrm{DMode}+0.03\mathrm{GAge}+0.03\mathrm{MAge}+0.86\mathrm{SES}_2+1.08+0.72\mathrm{SES}_4+0.81\mathrm{SES}_5-0.575\mathrm{MSmk}+log(OR_{A-U})\,\mathrm{U}))$<br><br>where $\mathrm{Eth}_2, \mathrm{Eth}_3$ denotes the indicator variables for Asian and other, while $\mathrm{Sibs}_2, \mathrm{Sibs}_3, \mathrm{Sibs}_4$ denotes indicator variables for two, three and four siblings<br><br>Model parameters between exposure and observed confounders were based on associations observed in the real case study data using the mismeasured exposure and ignoring the unmeasured confounder, while associations between exposure and the unmeasured confounder ($OR_{A-U}$) were set to values sourced from literature confirmed by substantive experts (see Table 3 in main text)<br><br>*<u>Selection- type II due to restricting analysis to those who can speak/understand English</u>*<br><br>8. Ability to speak/understand English (yes/no) was generated from a binomial distribution using probability sourced from the literature (see Table 3 in main text)<br>$\mathrm{E}\sim Bernoulli(logit^{-1}(p_E))$<br>*<u>Outcome (true unobserved)</u>*<br><br>9. Finally, the outcome was generated using a logistic regression model conditioning on all confounders (including the unmeasured confounder) and the true exposure (generated in step 7) and an interaction between exposure and ability to speak/understand English<br><br>$\mathrm{Y}\sim Bernoulli(logit^{-1}(-2.39+\log(\alpha)\,\mathrm{A}-0.63\mathrm{Sex}+0.05\mathrm{Sibs}_2+0.02\mathrm{Sibs}_3-0.02\mathrm{Sibs}_4+0.19\mathrm{LBW}-0.04\mathrm{PEth}_2+0.07\mathrm{PEth}_3+0.95\mathrm{FMa}+0.69\mathrm{FPa}-0.11\mathrm{FHx}+0.17\mathrm{DMode}+0.19\mathrm{GAge}+0.02\mathrm{MAge}+0.11\mathrm{SES}_2-0.05\mathrm{SES}_3-0.12\mathrm{SES}_4-0.02\mathrm{SES}_5+0.80\mathrm{MSmok}+\log(OR_{Y-U})\mathrm{U}+0.18\mathrm{E}+\log(OR_{(Y-A)E})\,\mathrm{A}\times\mathrm{E})$<br><br>Model parameters between outcome and observed confounders were closely based on associations between observed in the real case study data using the mismeasured outcome and exposure, and ignoring the unmeasured confounder, missing data and effect modifier, while association between outcome and the unmeasured confounder ($OR_{Y-U}$) and regression coefficient for the interaction term were elicited via expert opinion (see Table 3 in main text). The regression coefficient for exposure ($\alpha$)was set to a fixed value determined via trial and error to achieve ~ 80% power for estimating the ACE using the |

target analysis models at the conventional 0.05 significance level in all scenarios (before introducing any missing values). Based on the above criterion, under "realistic" scenario, $\alpha = (-0.43)$, under "enhanced" scenario, $\alpha$= (-0.05)

*Misclassified exposure and outcome*

10. Misclassified exposure was generated using a logistic regression model conditional on the true exposure generated in step 7 and the observed confounders generated in steps 1-5. Model parameters between misclassified exposure and true exposure ($OR_{A^*-A}$)were sourced from literature confirmed by substantive experts (see Table 3 in main text) and those between the misclassified exposure and observed confounders were closely based on associations in the case study.

$$A^* \sim Bernoulli(logit^{-1}(1.79 - 0.17Sex - 0.56NSibs_2 - 0.99NSibs_3 - 1.06NSibs_4 - 0.38LBW + 0.31PEth_2 + 0.36PEth_3 + 0.04FMa - 0.11FPa - 0.01FHx - 0.10DMode + 0.03GAge + 0.03MAge + 0.86SES_2 + 1.08 + 0.72SES_4 + 0.81SES_5 - 0.575MSmk + \log(OR_{A^*-A})A$$

Where $OR_{A^*-A} = (TP/FN)/(FP/TN)$ and $\mathrm{TP} = \mathrm{p_A} \times \mathrm{N} \times sens, FN = (p_A \times N) - TP, \mathrm{TN} = (1 - \mathrm{p_A}) \times \mathrm{N} \times spec$ $, FP = \left((1 - p_A) \times N\right) - TN$

$$\mathrm{p_A} = \text{prevalence of A in the population} = 0.94$$

$TP$: true positives , $TN$: true negatives, $FN$: false negatives, $FP$: false positives, $sens$: sensitivity, $spec$: specificity.

11. Misclassified outcome was generated using a logistic regression model conditional on true outcome and the observed confounders with model parameters between the misclassified outcome and the true outcome ($OR_{Y^*-Y}$) sourced from literature confirmed by substantive experts (see Table 3 in main text) and those between the misclassified outcome and observed confounders based on associations in the case study.

$$Y^* \sim Bernoulli(logit^{-1}(-2.39 - 0.63Sex + 0.05Sibs_2 + 0.02Sibs_3 - 0.02Sibs_4 + 0.19LBW - 0.04PEth_2 + 0.07PEth_3 + 0.95FMa + 0.69FPa - 0.11FHx + 0.17DMode + 0.19GAge + 0.02MAge + 0.11SES_2 - 0.05SES_3 - 0.12SES_4 - 0.02SES_5 + 0.80MSmok + \log(OR_{Y^*-Y})Y$$

Where $OR_{Y^*-Y} = (TP/FN)/(FP/TN)$ and $TP = p_Y \times N \times sens, FN = (p_Y \times N) - TP, TN = (1 - p_Y) \times N \times spec, FP = \left((1 - p_Y) \times N\right) - TN$

$$p_Y = \text{prevalence of Y in the population} = 0.10$$

| | |
|---|---|
| | $TP$: true positives, $TN$: true negatives, $FN$: false negatives, $FP$: false positives, $sens$: sensitivity, $spec$: specificity.<br><br>*Selection- type I missingness in outcome*<br><br>12. Missingness in the outcome (i.e., an indicator variable for indicating missing or observed) was generated using a logistic regression model conditional on true exposure, true outcome and observed confounders. Model parameters for exposure and confounders were set to values close to those in observed data (using mismeasured exposure and ignoring outcome). The regression coefficient for the outcome, $OR_{Y-R_y}$, were elicited via expert opinion.<br><br>$R_Y \sim Bernoulli(logit^{-1}(-1.74 + 0.79A - 0.11Sex - 0.19Sibs_2 - 0.30Sibs_3 - 0.52Sibs_4 - 0.18LBW - 0.29Eth_2 - 0.35PEth_3 + 0.22FMa + 0.21FPa + 0.29FH - 0.13DMode + 0.13GAge + 0.06MAge + 0.07SES_2 + 0.21SES_3 + 0.13SES_4 - 0.01SES_5 - 0.54MSmok + \log(OR_{Y-R_y})Y$<br><br>**Factors that were varied:**<br><br>The strengths of the biases were varied by varying the magnitude of the following parameters:<br>• Selection bias (collider stratification bias): $OR_{Y-R_y}$<br>• Selection bias (generalizability): $(p_E)$, $OR_{(Y-A)E}$<br>• Confounding bias: $p_U$, $OR_{A-U}$, $OR_{Y-U}$<br>• Exposure misclassification bias: $OR_{A^*-A}$<br>• Outcome misclassification bias: $OR_{Y^*-Y}$<br><br>One "realistic value" was picked based on the literature and expert opinion (realistic scenario), and then varied ×2 (enhanced scenario). See main text Table 3 for the values. All bias parameters were varied simultaneously in each scenario. |
| **Estimand/target of analysis** | • The estimand of interest is the ACE of breast-feeding at least once vs never on asthma (yes/no) at age 6. Two casual effect measures were considered: risk difference ($\beta_{RD}$ )and risk ratio ($\beta_{RR}$) |
| | • Each dataset was analysed using the models in (1), which are misspecified under the data generation. Thus, to avoid our results being affected by this misspecification bias, as "true" value for computing bias we used estimates for the exposure coefficient obtained by fitting these models to a very large sample. |
| **Methods** | For both (i) and (ii) below, the multiple bias adjustment method via imputation by Brendel et al [3] was used:<br><br>(i)one-at- time bias adjustment, producing multiple bias-adjusted estimates.<br>(ii)simultaneous adjustment of multiple biases producing a single bias-adjusted estimate<br><br>To assess the robustness of the analytic strategies to misspecification of bias parameters, analyses will be repeated using:<br><br>• The correct bias parameters<br>• The correct bias parameters multiplied by a factor of 2 |
| **Performance measures** | Bias: Average difference between the bias-adjusted estimate and the true value<br>Relative bias: Bias relative to the true value as a percentage<br>Empirical SE: Standard deviation of the bias adjusted estimate<br>Model-based SE: Average of the estimated standard error of the bias-adjusted estimate<br>Coverage probability: Proportion of replications where the estimated 95% confidence interval contained the true value<br>Bias-eliminated coverage: Proportion of replications in which the estimated 95% CI contained the mean value of the bias-adjusted estimate (across replications)<br><br>Note : Under-coverage could be due to bias and/or model-based SE$\neq$ empirical SE. Bias-eliminated coverage, eliminates the bias of a method from confidence intervals by studying confidence interval coverage for mean of $\hat{\beta}_{RD}$ and $\hat{\beta}_{RR}$. [1] |

*$logit^{-1}\big((.)\big) = \exp(.)\,/(1+\exp(.))$

Table S2: Bias parameter distributions used in the case study application

| Bias | Bias parameter | Source(s) | Estimate | Conditional bias parameter distributions used in the case study |
|---|---|---|---|---|
| Confounding bias | Odds of gestational hypertension | Fox, Haylee, and Emily J. Callander. "The cost of hypertensive disorders of pregnancy to the Australian healthcare system." Pregnancy Hypertension 21 (2020): 197-199. https://pubmed.ncbi.nlm.nih.gov/32634609/ | 0.03 | Normal distribution deduced by specifying 0.07 and 0.17 as the 2.5$^{th}$ and 97.5$^{th}$ percentiles and 0.10 as the mean[4] $N(\log(0.12), \frac{\log(0.20) - \log(0.07)}{2 \times 1.96})$ |
| | | Arnott, Clare, et al. "Maternal cardiovascular risk after hypertensive disorder of pregnancy." Heart 106.24 (2020): 1927-1933. https://heart.bmj.com/content/106/24/1927 | 0.11 | |
| | | Wertaschnigg, Dagmar, et al. "Hypertensive disorders in pregnancy–Trends over eight years: A population-based cohort study." *Pregnancy hypertension* 28 (2022): 60-65. https://www.sciencedirect.com/science/article/abs/pii/S2210778922000277 | 0.06 | |
| | | Shaheen, Seif O., et al. "Hypertensive disorders of pregnancy, respiratory outcomes and atopy in childhood." European Respiratory Journal 47.1 (2016): 156-165. | 0.17 | |
| | | https://erj.ersjournals.com/content/erj/47/1/156.full.pdf | | |
| Confounding bias | Odds ratio for the association between ever being breastfed and gestational hypertension | Horsley, Kristin, et al. "Hypertensive disorders of pregnancy and breastfeeding practices: A secondary analysis of data from the All Our Families Cohort." Acta Obstetricia et Gynecologica Scandinavica 101.8 (2022): 871-879. https://obgyn.onlinelibrary.wiley.com/doi/epdf/10.1111/aogs.14378 | 0.45 | $N\left(\log(0.70), \frac{\log(0.80)-\log(0.50)}{2\times 1.96}\right)$ |
| | | Magnus, Maria C., et al. "Breastfeeding and later-life cardiometabolic health in women with and without hypertensive disorders of pregnancy." Journal of the American Heart Association 12.5 (2023): e026696. https://www.ahajournals.org/doi/epub/10.1161/JAHA.122.026696 | 0.76 | |
| | | Leeners B, Rath W, Kuse S, Neumaier-Wagner P. Breast-feeding in women with hypertensive disorders in pregnancy. J Perinat Med. 2005;33(6):553-60. doi: 10.1515/JPM.2005.099. PMID: 16318622. https://www.degruyter.com/document/doi/10.1515/JPM.2005.099/html | 0.67 | |
| | | Woolley, Emma, et al. "Exclusive Breastfeeding at Discharge in Regional New South Wales, Australia: The Role of Antenatal Care (2011–2020)." International Journal of Environmental Research and Public Health 20.12 (2023): 6135. https://www.mdpi.com/1660-4601/20/12/6135 | 0.60 | |
| Confounding bias | Odds ratio for the association between asthma and gestational hypertension | Henderson, Ian, and Siobhan Quenby. "Gestational hypertension and childhood atopy: a Millennium Cohort Study analysis." European Journal of Pediatrics 180 (2021): 2419-2427. https://link.springer.com/article/10.1007/s00431-021-04012-3 | 1.25 | $N(\log(1.1)\,, \frac{\log(0.88) - \log(1.2)}{2 \times 1.96})$ |
| | | Shaheen, Seif O., et al. "Hypertensive disorders of pregnancy, respiratory outcomes and atopy in childhood." European Respiratory Journal 47.1 (2016): 156-165. https://erj.ersjournals.com/content/erj/47/1/156.full.pdf | 1.03 | |
| | | Kelly, L., et al. "OP99 Hypertensive disorders in pregnancy and childhood diagnosis of asthma." (2019): A48-A49. https://jech.bmj.com/content/jech/73/Suppl_1/A48.2.full.pdf | 1.37 | |
| | | Ma, Ying, et al. "Associations between maternal complications during pregnancy and childhood asthma: a retrospective cohort study." ERJ Open Research 9.2 (2023). https://openres.ersjournals.com/content/erjor/9/2/00548-2022.full.pdf | 1.65 | |
| Measurement bias- breastfeeding self-reports | Sensitivity | Li, Ruowei, Kelley S. Scanlon, and Mary K. Serdula. "The validity and reliability of maternal recall of breastfeeding practice." Nutrition reviews 63.4 (2005): 103-110. The validity and reliability of maternal recall of breastfeeding practice - PubMed (nih.gov) | 0.82 | $beta(44,4)$ Beta distribution deduced by specifying 0.82 and 0.90 as the 2.5$^{th}$ and 97.5$^{th}$ percentiles and 0.84 as the prior mean[4] |
| | | Amissah, Emma Ayorkor, et al. "Validation study of maternal recall on breastfeeding duration 6 years after childbirth." Journal of Human Lactation 33.2 (2017): 390-400. https://www.ncbi.nlm.nih.gov/pmc/articles/PMC9353757/ | 0.91 (0.88-0.94) | |
| | | Schneider, Bruna Celestino, et al. "Validation of maternal recall on exclusive breastfeeding 12 months after childbirth." Public health nutrition 23.14 (2020): 2494-2500. | 0.98 (0.97-0.99) | |
| | | Validation of maternal recall on exclusive breastfeeding 12 months after childbirth - PMC (nih.gov) | | |
| Measurement bias- breastfeeding self-reports | Specificity | Li, Ruowei, Kelley S. Scanlon, and Mary K. Serdula. "The validity and reliability of maternal recall of breastfeeding practice." Nutrition reviews 63.4 (2005): 103-110. The validity and reliability of maternal recall of breastfeeding practice - PubMed (nih.gov) | 0.93 | $beta(82,18)$ Beta distribution deduced by specifying 0.70 and 0.85 as the 2.5$^{th}$ and 97.5$^{th}$ percentiles and 0.84 as the prior mean[4] |
| | | Amissah, Emma Ayorkor, et al. "Validation study of maternal recall on breastfeeding duration 6 years after childbirth." Journal of Human Lactation 33.2 (2017): 390-400. https://www.ncbi.nlm.nih.gov/pmc/articles/PMC9353757/ | 0.89 (0.84-0.92) | |
| | | Schneider, Bruna Celestino, et al. "Validation of maternal recall on exclusive breastfeeding 12 months after childbirth." Public health nutrition 23.14 (2020): 2494-2500. Validation of maternal recall on exclusive breastfeeding 12 months after childbirth - PMC (nih.gov) | 0.70 (0.68-0.72) | |
| Measurement bias- | True population prevalence | Ogbo, Felix A., et al. "Prevalence and determinants of cessation of exclusive breastfeeding in the early | 0.90 | $U(0.80,0.90)$ |
| breastfeeding self-reports | | postnatal period in Sydney, Australia." *International Breastfeeding Journal* 12 (2016): 1-10. https://link.springer.com/article/10.1186/s13006-017-0110-4 | | *note true population prevalence assumed to lower than values from literature based on self-reports |
| | | Netting, Merryn J., et al. "The Australian feeding infants and toddler study (OzFITS 2021): Breastfeeding and early feeding practices." *Nutrients* 14.1 (2022): 206. https://www.mdpi.com/2072-6643/14/1/206 | 0.93 | |
| | | Scott, Jane, et al. "Determinants of continued breastfeeding at 12 and 24 months: results of an Australian cohort study." *International journal of environmental research and public health* 16.20 (2019): 3980. https://www.mdpi.com/1660-4601/16/20/3980 | 0.95 | |
| Measurement bias- Child asthma parent-reports | Sensitivity | Hederos, Carl-Axel, et al. "Comparison of clinically diagnosed asthma with parental assessment of children's asthma in a questionnaire." Pediatric allergy and immunology 18.2 (2007): 135-141. | 0.85 | $beta(54,13)$ Beta distribution deduced by specifying 0.77 and 0.90 as the 2.5$^{th}$ and 97.5$^{th}$ percentiles and 0.80 as the prior mean[4] |
| | | Jenkins, Mark A., et al. "Validation of questionnaire and bronchial hyperresponsiveness against | 0.77 | |
| | | respiratory physician assessment in the diagnosis of asthma." International journal of epidemiology 25.3 (1996): 609-616. | | |
| | | Valle, Solange Oliveira Rodrigues, et al. "Validity and reproducibility of the asthma core International Study of Asthma and Allergies in Childhood (ISAAC) written questionnaire obtained by telephone survey." Journal of Asthma 49.4 (2012): 390-394. | 0.87 (0.78-0.96) | |
| Measurement bias- Child asthma parent-reports | Specificity | Hederos, Carl-Axel, et al. "Comparison of clinically diagnosed asthma with parental assessment of children's asthma in a questionnaire." Pediatric allergy and immunology 18.2 (2007): 135-141. | 0.81 | $beta(42,5)$<br>Beta distribution deduced by specifying 0.81 and 0.95 as the 2.5$^{th}$ and 97.5$^{th}$ percentiles and 0.90 as the prior mean[4] |
| | | Jenkins, Mark A., et al. "Validation of questionnaire and bronchial hyperresponsiveness against respiratory physician assessment in the diagnosis of a sthma." International journal of epidemiology 25.3 (1996): 609-616. | 0.98 | |
| | | Valle, Solange Oliveira Rodrigues, et al. "Validity and reproducibility of the asthma core International Study of Asthma and Allergies in Childhood | 1.00 (0.90-1.00) | |
| | | (ISAAC) written questionnaire obtained by telephone survey." Journal of Asthma 49.4 (2012): 390-394. | | |
| Measurement bias- Child asthma parent-reports | Population prevalence | Khan, J.R., Lingam, R., Owens, L. et al. Social deprivation and spatial clustering of childhood asthma in Australia. glob health res policy 9, 22 (2024). https://doi.org/10.1186/s41256-024-00361-2 | 0.21 | $U(0.21,0.30)$ |
| | | Ahmad, K., Kabir, E., Ormsby, G.M. *et al.* Are wheezing, asthma and eczema in children associated with mother's health during pregnancy? Evidence from an Australian birth cohort. *Arch Public Health* **79**, 193 (2021). https://doi.org/10.1186/s13690-021-00718-w | 0.26 | |
| | | Thomson, Jennifer A., et al. "Early childhood infections and immunisation and the development of allergic disease in particular asthma in a high-risk cohort: a prospective study of allergy-prone children from birth to six years." *Pediatric allergy and immunology* 21.7 (2010): 1076-1085. https://onlinelibrary.wiley.com/doi/full/10.1111/j.1399-3038.2010.01018.x | 0.25(0.22-0.30) | |
| Measurement bias- Child asthma parent-reports | Association between true asthma and true breastfeeding ^ | Lodge, Caroline J., et al. "Breastfeeding and asthma and allergies: a systematic review and meta-analysis." Acta paediatrica 104 (2015): 38-53. https://doi.org/10.1111/apa.13132 | | $N(\log(0.6), \frac{\log(0.8) - \log(0.5)}{2 \times 1.96})$ |
| Selection bias due to restricting study sample to those who consented | Odds of consent | Koplin, Jennifer J., et al. "Cohort Profile: The HealthNuts Study: Population prevalence and environmental/genetic predictors of food allergy." International journal of epidemiology 44.4 (2015): 1161-1171. https://doi.org/10.1093/ije/dyu261<br><br>and expert opinion | 2.3-4 (prevalence 0.70-0.8) | $N(\log(2.8), \frac{\log(4) - \log(2.3)}{2 \times 1.96})$ |
| | Odds ratio for the association between consent and breastfeeding | Elicited via expert opinion | | $N(\log(1.1), \frac{\log(1.3) - \log(1.0)}{2 \times 1.96})$ |
| | Odds ratio for the association between consent and asthma | Elicited via expert opinion | | $N(\log(1.2), \frac{\log(1.3) - \log(0.9)}{2 \times 1.96})$ |
| | Effect measure modification of consent -asthma by breastfeeding (Relative change in OR of consent -asthma in those who were ever breast-fed vs not) | Elicited via expert opinion | | $N(\log(1.4), \frac{\log(1.7) - \log(1.0)}{2 \times 1.96})$ |
| Selection bias due to restricting study sample | Odds of ability to speak and understand English in Victoria from 2016-2021 | https://profile.id.com.au/australia/speaks-english | 4-9 (under prevalence 0.80-0.90) | $N(\log(5), \frac{\log(9) - \log(4)}{2 \times 1.96})$ |
| to those who read/understand English | | | | |
| | Odds ratio for the association between ability to read/understand English and breast feeding | Dennis, Cindy-Lee, et al. "Breastfeeding rates in immigrant and non-immigrant women: A systematic review and meta-analysis." Maternal & Child Nutrition 15.3 (2019): e12809. https://onlinelibrary.wiley.com/doi/full/10.1111/mcn.12809 | 0.88(0.84-0.93) | $N(\log(0.88),\frac{\log(0.93)-\log(0.84)}{2\times 1.96})$ |
| | Odds ratio for the association between ability to read/understand English and asthma | Wang, Yichao, et al. "Asian children living in Australia have a different profile of allergy and anaphylaxis than Australian-born children: a State-wide survey." *Clinical & Experimental Allergy* 48.10 (2018): 1317-1324. https://doi.org/10.1111/cea.13235 and expert opinion | | $N(\log(1.2),\frac{\log(1.5)-\log(0.90)}{2\times 1.96})$ |
| | Effect measure modification of ability to read/understand English -asthma by breastfeeding (Relative change in OR of ability to speak English-asthma in those who were ever breast-fed vs not) | Elicited via expert opinion | | $N(\log(0.70),\frac{\log(0.80)-\log(0.5)}{2\times 1.96})$ |
| Selection bias due to missing | Odds of response for | Elicited via expert opinion | | $N(\log(1.2),\frac{\log(1.5)-\log(1.10)}{2\times 1.96})$ |

| | | |
|---|---|---|
| data in the outcome | parents/guardians with asthmatic children vs parents/guardians with non- asthmatic children | |

* Conditional bias parameters were often approximated using marginal bias parameter values in the literature. Systematic approaches for calibrating marginal bias parameters to conditional bias parameters within this bias adjustment approach is still an area of further research.

^ Bias parameter only required for simultaneous bias adjustment

**Note:** Conditional bias parameter distributions for the observed confounder variables were estimated by fitting a logistic regression model for the corresponding observed variable (i.e. $Y^*$ when imputing $Y$, and $A^*$ when imputing $A$) or the selection indicators with confounders as predictors

Table S3: Application of the bias-adjustment strategies to the HealthNuts case study

| Approach | Biases adjusted for in bias analysis | Causal effect measure | | | |
|---|---|---|---|---|---|
| | | **Risk difference** | **95% SI** | **Risk ratio** | **95% SI** |
| Primary analysis | NA | -0.05 | -0.11, 0.02 * | 0.97 | 0.82,1.15 * |
| Simultaneous bias adjustment | All biases | -0.10 | -0.26, 0.05 | 0.77 | 0.54 ,1.14 |
| One-at-a-time bias adjustments | SB-collider stratification | -0.03 | -0.09, 0.03 | 0.82 | 0.56 ,1.37 |
| | SB-generalizability (consent) | -0.03 | -0.1, 0.03 | 0.79 | 0.54,1.31 |
| | SB-generalizability (English speaking/ethnicity) | -0.04 | -0.1,0.03 | 0.77 | 0.53 ,1.29 |
| | MB-A | -0.05 | -0.19, 0.07 | 0.73 | 0.41 ,1.92 |
| | MB-Y | -0.07 | -0.19, 0.06 | 0.86 | 0.64,1.14 |
| | CB | -0.03 | -0.10, 0.04 | 0.81 | 0.57, 1.37 |

*95% CI shown

NA: Not applicable, CB: Confounding bias, MB: measurement bias, SB: Selection bias, SI: simulation interval